\documentclass[preprint,11pt]{aastex}
\usepackage{epsfig}
\usepackage{times}
\shorttitle{SiO in the HH~211 jet}
\shortauthors{Chandler \& Richer}

\begin{document}

\title{Sub-arcsecond Imaging of SiO in the HH~211 Protostellar Jet}
\author{Claire J. Chandler}
\affil{National Radio Astronomy Observatory, PO Box O, Socorro, NM
87801, USA}
\and
\author{John S. Richer}
\affil{Mullard Radio Astronomy Observatory, Cavendish Laboratory,
Madingley Road, \\ Cambridge CB3 0HE, United Kingdom}
\email{cchandle@nrao.edu, jsr@mrao.cam.ac.uk}

\begin{abstract}

We present images of the HH~211 molecular jet in the SiO $v$=0, $J$=1--0
line at 43~GHz made with the Very Large Array at approximately 0.5
arcsec resolution.  The SiO emission appears to trace primarily internal
bowshocks in the outflow, suggesting that the dust and molecular gas
are accelerated via prompt entrainment at internal working surfaces in
the jet.  There is also some evidence for limb-brightening of the SiO
emission, indicating that SiO emission may also arise from entrainment in
the jet's boundary layer.  Excitation temperatures of $\ga 150$--200~K
are inferred from the SiO emission.  Enhancements in the SiO abundance
of $\sim 10^6$ over interstellar values are observed, and the possible
origin of the SiO is discussed.

\end{abstract}


\section{Introduction}

Ever since protostellar outflows were first discovered our understanding
of their driving mechanism has been limited by poor spatial resolution.
As a consequence, early measurements (e.g., Snell, Loren, \& Plambeck
1980; Lada 1985) suggested the flows had relatively small length-to-width
ratios, and the first models concentrated on explaining high-velocity
molecular gas as being driven by wind-angle protostellar winds (e.g.,
Barral \& Cant\'o 1981; K\"onigl 1982; Boss 1987; Shu et al.\ 1991).
This ruled out on morphological grounds the possibility that Herbig-Haro
(HH) jets, traced by low-excitation optical lines, were responsible
for the molecular flows.  However, more recent, higher resolution
molecular observations made with large single dishes and millimetre-wave
interferometers have demonstrated that the molecular flows frequently
possess highly-collimated components (e.g., Bachiller \& Cernicharo 1990),
and that these are often closely associated with HH objects (Richer et
al.\ 1989).  A large number of highly-collimated molecular outflows
have now been detected, and it is commonly assumed that these flows,
at least in the younger sources, are driven by the momentum of a high
Mach number, neutral jet (Richer et al.\ 2000).

The detailed physics of how neutral jets, with densities comparable to
that of the surrounding material, accelerate molecular gas to generate the
molecular flow remain unclear, although several models have been proposed.
Material can in principle be entrained in the leading bowshock (e.g.,
Raga \& Cabrit 1993), internal bowshocks (e.g., Raga et al.\ 1993), and
in a turbulent mixing layer along the edges of the jet (Stahler 1994;
Taylor \& Raga 1995).  In most cases, however, the angular resolution
of current observations is insufficient to resolve the details of these
processes.  What is needed is a shock tracer similar to the molecular
hydrogen lines at 2~$\mu$m, but which does not suffer the obscuration
prevalent at optical and near-infrared wavelengths.

The SiO molecule is a very useful tool for investigating jet-cloud
interactions (e.g., Mart\'\i n-Pintado, Bachiller, \& Fuente 1992).
It can be observed at high angular resolution in the millimetre wavebands,
and it seems to be excited primarily in shocks where silicon is removed
from dust grains, significantly increasing its gas phase abundance.
Thermal emission from its vibrational ground state lines has been
detected in a number of protostellar outflow sources, and it appears to
be intimately associated with collimated jets (e.g., L1448: Guilloteau
et al.\ 1992) rather than with the general, low-velocity outflow traced
by CO emission.  Models suggest that SiO most likely originates from
the sputtering of silicon atoms off silicate grains in C-shocks having
speeds in the range 10 to 50~km~s$^{-1}$, followed by rapid conversion to
SiO in the gas phase (Field et al.\ 1997; Schilke et al.\ 1997; Caselli,
Hartquist, \& Havnes 1997).  Abundance enhancements of 4 to 5 orders of
magnitude over the very low gas-phase abundance of SiO in the ambient
interstellar medium have been observed in some outflows (e.g., Bachiller,
Mart\'\i n-Pintado, \& Fuente 1991; Mikami et al.\ 1992; Bachiller et
al.\ 1994).  The lack of foreground extinction at millimetre wavelengths
also make the rotational transitions of SiO particularly important for
studies of the youngest outflows, which are associated with the most
deeply embedded protostars.

The protostar driving the HH~211 system is one of the youngest and
most deeply embedded low-mass protostars known.  It was discovered
serendipitously through its bright, shocked, H$_2$ 2-$\mu$m emission in
an imaging study of the IC348 region (McCaughrean, Rayner, \& Zinnecker
1994), and lies at a distance of $\sim 350$~pc (Herbig \& Jones 1983).
Although given a Herbig-Haro catalogue number, no {\it optical} emission
has yet been detected from this object, most likely due to the high
obscuration.  The H$_2$ image shows a compact, well-collimated flow which
is also detected in CO emission (Gueth \& Guilloteau 1999).  Because of
the relatively small projected CO flow speeds ($<20$~km~s$^{-1}$), and
the high degree of separation of the red and blue lobes, the outflow's
inclination to the plane of the sky is probably small; if the true CO
flow speeds are $>100$~km~s$^{-1}$ as in other outflows, the implied
inclination is less than $10^\circ$.  Between the two lobes, the driving
protostar is detected as a millimetre continuum source with no 2-$\mu$m or
optical counterpart, designated HH~211-mm by Gueth \& Guilloteau (1999).
The outflow's age, and hence that of the protostar, can be estimated from
the outflow properties, although various assumptions need to be made.
The value derived by Gueth \& Guilloteau (1999) is only 750 years,
suggesting that the embedded protostar has only just entered its main
accretion and outflow phase, and making it an ideal object in which to
study the early stages of low mass star formation.  Here we have taken
advantage of the high resolution available with the Very Large Array
(VLA) at 43~GHz to study the SiO $v$=0, $J$=1--0 emission from the HH~211
protostellar jet, and to investigate the entrainment and acceleration
of molecular outflows.

\section{Observations and data reduction}

Observations of HH~211 in the $v$=0, $J$=1--0 transition of SiO at
43.4~GHz were made using 10 antennas of the VLA of the National Radio
Astronomy Observatory\footnote{The National Radio Astronomy Observatory is
a facility of the National Science Foundation operated under cooperative
agreement by Associated Universities, Inc.}.  The FWHM of the primary
beam of the VLA antennas at this frequency is approximately 1 arcmin.
The first observations took place in the compact D configuration, in 1996
August; we re-observed HH~211 using the more extended C configuration in
1997 September to obtain higher resolution images.  On each occasion,
two circular polarisations were measured each with 16 Hanning-smoothed
channels, providing a spectral resolution of 781.25~kHz (5.4~km~s$^{-1}$).

In D configuration two phase centers were observed, one toward the
red outflow lobe and one toward the blue.  The positions are listed in
Table~\ref{vlatab}.  Atmospheric and instrumental phase fluctuations
were monitored every 10 minutes by observing the quasar Pks~0333+321,
and the pointing was corrected using 8.4~GHz observations of this same
quasar at intervals of approximately one hour.  The bandpass response was
obtained from observations of 3C84.  The absolute flux density scale was
determined from 3C48, with an assumed 43~GHz flux density of 0.57~Jy, to
give a measured 43~GHz flux density for 0333+321 of $1.40 \pm 0.03$~Jy.
The remaining 14 VLA antennas were used to search for 22.2~GHz H$_2$O
maser emission within the primary beam of 2 arcmin (FWHM).  A single
IF with right circular polarisation was used with 64 Hanning-smoothed
channels covering a total bandwidth of 6.25~MHz (84.3~km~s$^{-1}$),
centered on $V_{\rm LSR} = 0$~km~s$^{-1}$.  A single phase center
was observed, and the same quasar was used at 22.2~GHz for the phase,
bandpass, and flux density calibration as at 43.4~GHz.  The synthesised
beam was 3\farcs1~$\times$~2\farcs5 at PA = 73$^\circ$, and the rms
noise per channel was 9~mJy per beam.  No H$_2$O emission was detected
from HH~211.

The C-array observations used a single phase center, at the location of
the protostar.  The atmospheric phase fluctuations on the longer baselines
in C configuration are significantly worse than in D, so we used a much
faster switching time between the source and 0333+321, with a total cycle
time from source to calibrator to source of 1 min 40 sec.  The pointing
was checked every hour using 8.4~GHz observations of 0333+321, and the
bandpass was obtained from 3C84.  The absolute flux density scale was
determined from 3C286, with an assumed 43~GHz flux density of 1.45~Jy,
to give a 43~GHz flux density for 0333+321 of $2.08 \pm 0.08$~Jy on
this occasion.  The uncertainty in the absolute flux density scale for
both datasets is approximately 15\%.

The data were calibrated, imaged, and CLEANed using the AIPS reduction
package, and the synthesised beam, which is intermediate between uniform
and natural weighting (``robust'' weighting: Briggs 1995), is optimised
for a combination of resolution and sensitivity.  The CLEAN beam widths
and rms noise per channel at the center of each field are listed in
Table~\ref{vlatab}; however, because the two fields in D configuration
were corrected for the response of the primary beam and coadded, they
have both been restored with the average synthesised beam listed in
column 5.  All the quantitative results described below are derived from
CLEANed images of the data obtained in each VLA configuration separately.
However, images have also been made by combining the C and D-array data
and carrying out a joint deconvolution of the synthesized beams using
the maximum entropy routine UTESS in AIPS\@.  These images are displayed
in Figs~\ref{sio-int} and \ref{sio-chan}, but because the resolution
is a function of the local signal-to-noise ratio after maximum-entropy
processing, the UTESS images have not been used in any of the quantitative
analyses.

\section{Results and discussion}

\subsection{Structure and kinematics}

Fig.~\ref{sio-int} shows the integrated red and blueshifted SiO(1--0)
emission from HH~211 overlaid on a logarthmic greyscale image of the
shocked molecular hydrogen emission from McCaughrean et al.\ (1994).
The SiO is mainly located inside the outer faint shell of H$_2$ emission,
is well collimated, and appears to be associated only with the jet from
the exciting star.  It corresponds closely to the strongest emission
peaks in the high-velocity CO(2--1) map of Gueth \& Guilloteau (1999).
While the SiO emission on the jet axis is generally coincident with
regions of shocked H$_2$ (in the case of the redshifted component,
even terminating at an H$_2$ knot), no emission is detected at the
brightest H$_2$ peaks despite the fact that they fall well inside the
50\% response level of the VLA D-array observations (dashed contour,
Fig.~\ref{sio-int}).  This lack of detailed correlation between the two
shock tracers is also seen in other young outflows (e.g., IRAS~03828+3035:
Bachiller et al.\ 1994; L1157: Gueth, Guilloteau, \& Bachiller 1998),
and possible explanations are discussed further in \S~\ref{excite} below.

Between the position of the protostar and a projected angular distance of
15$''$ both the red and blue components of the SiO jet gradually widen.
Little or no redshifted SiO is detected beyond 15$''$, although knots and
other features continue out to larger distances on the blueshifted side.
The total flux in the blueshifted and redshifted SiO jets of HH~211
is comparable, but Fig.~\ref{sio-int} shows that the blue emission
is more diffuse than the red, a feature which is illustrated quite
dramatically in maps of the emission made within individual velocity
channels (Fig.~\ref{sio-chan}).  The systemic velocity, $V_{\rm
sys}$, of the molecular cloud is at $V_{\rm LSR} \sim 9$~km~s$^{-1}$
(Gueth \& Guilloteau 1999), and at a radial velocity $|V-V_{\rm sys}|
\sim 10$~km~s$^{-1}$ the full width zero intensity of the blueshifted
emission is $\sim 4''$, twice that on the redshifted side.  A further
difference between the structures of the red and blue SiO jets is in
their association with infrared H$_2$ emission: the one strong peak in the
blueshifted jet, at $V_{\rm LSR} = -0.8$~km~s$^{-1}$, corresponds closely
to a bright H$_2$ knot, while the redshifted SiO is coincident only with
very faint H$_2$ emission.  These features are most easily explained
by the structure of the molecular cloud core in which HH~211 resides.
The core exhibits a steep density gradient in the direction of the
blueshifted side of the jet, compared with the considerable amounts
of dense gas on the redshifted side (traced by H$^{13}$CO$^+$(1--0)
emission: see Gueth \& Guilloteau 1999) into which the redshifted
jet must penetrate.  The decreased pressure of the ambient gas on the
blueshifted side may allow the jet to expand more freely perpendicular to
the jet direction.  Toward the redshifted SiO jet the dense gas results
in high foreground extinction for the infrared H$_2$ emission.

In three of the redshifted channels, from $V_{\rm LSR}$ = 20.8 to
31.6~km~s$^{-1}$, the SiO emission divides into two streamers separated
by $\sim 1''$, and appears to be the limb-brightening of a layer of
molecular material around the faint H$_2$ jet.  At several locations the
bifurcated streamers seem to lie behind (i.e., closer to the protostar)
a knot of SiO emission on the jet axis that is at a higher velocity.
Such a configuration suggests that the SiO is accelerated at internal
bowshocks within the jet, and, indeed, may be formed in these bowshocks.
Alternatively, the limb brightening may be the result of laminar
entrainment of SiO along the edges of the jet.

The kinematics of the SiO jet matches closely that of the high-velocity
CO(2--1) detected by Gueth \& Guilloteau (1999).  The SiO emission
in the images made from only C-array data is very clumpy, and may be
missing flux particularly in the blueshifted channels.  We therefore
use only emission detected at a level of 3$\sigma$ or greater in the
VLA D configuration in analyzing the overall velocity structure of the
SiO jet.  Fig.~\ref{mom-D}a shows the first moment (intensity-weighted
mean velocity) of the SiO(1--0), overlaid with contours of shocked
H$_2$ emission.  It illustrates that the ends of the jets are the
locations of highest radial velocity compared to the ambient cloud.
An approximately uniform velocity gradient along the jet (sometimes called
a ``Hubble-law flow'') can also be seen, and the value of the gradient,
$\sim 5 \times 10^{-3}$~km~s$^{-1}$~AU$^{-1}$, is very similar to that
measured from the CO(2--1) emission.  The redshifted lobe in particular
shows a smooth velocity gradient along the jet, with a maximum at
30~km~s$^{-1}$ away from $V_{\rm sys}$.  The second moment map of the
SiO emission (Fig.~\ref{mom-D}b) shows that the positions of highest
velocity dispersion generally coincide with the brightest shocked H$_2$
emission on the jet axis, providing further evidence that the SiO is in
some way associated with, or possibly formed in, those jet shocks.

\subsection{Excitation, abundance, and origin of the SiO}
\label{excite}

The peak beam-averaged brightness temperature, $T_{\rm B}$, in the
C-array data is 54~K, and occurs in the channel at $V_{\rm LSR} =
26.2$~km~s$^{-1}$, providing a firm lower limit on the excitation
temperature, $T_{\rm ex}$, within the redshifted gas.  In the blueshifted
lobe the peak at $V_{\rm LSR} = -0.8$~km~s$^{-1}$ corresponds to
$T_{\rm B} = 47$~K\@.  Further constraints on $T_{\rm ex}$ can be
obtained by comparing the $J$=1--0 line with spectra of the SiO(5--4)
emission obtained with the James Clerk Maxwell Telescope in a 22$''$
beam (Chandler \& Richer 1997; Fig.~\ref{sio-spec}).  The integrated
$J$=1--0 flux from the C and D-array VLA images are the same to within
the calibration uncertainties, and the SiO jet is unresolved laterally
in the images made from data obtained only in the VLA D configuration.
We therefore assume that the D-array observations did not resolve out
any of the 1--0 emission.  If all the SiO is at the same temperature
and in local thermodynamic equilibrium both the SiO(5--4)/(1--0) line
ratio and the peak $T_{\rm B}$ can therefore be combined to give the
line optical depths and a lower limit to $T_{\rm ex}$.  We find for the
red jet that the $J$=1--0 optical depth, $\tau_{10}$, is $\sim 0.28$,
and $T_{\rm ex} \ga 220$~K, while for the blue jet $\tau_{10} \sim 0.36$
and $T_{\rm ex} \ga 155$~K\@.  These limits do not include uncertainties
in the flux calibration of the $J$=1--0 and 5--4 spectra, and assuming
a 15\% uncertainty in both gives strict lower limits to $T_{\rm ex}$ of
185~K (red) and 120~K (blue).  Such temperatures are considerably higher
than the $\sim 10$~K expected for the ambient molecular gas and dust.
Very high space densities in the emitting gas are also implied by the
detection of the SiO(5--4) line, which has a critical density of order
$10^6$~cm$^{-3}$.  These results confirm that the SiO originates in
regions of extreme excitation conditions.

The total mass of the high-velocity CO jet determined assuming a CO
abundance of 10$^{-4}$ is $\sim 2.5 \times 10^{-3} (T_{\rm ex}/50~{\rm
K})$~M$_\odot$ (Gueth \& Guilloteau 1999).  Assuming the SiO and
high-velocity CO(2--1) emission are co-spatial (but do not necessarily
have the same excitation temperature) a value $T_{\rm ex} = 150$~K for
the SiO gives an SiO abundance $X({\rm SiO}) \sim 6 \times 10^{-6}$,
comparable to that determined for the high-velocity gas in the L1448
jet by Mart\'\i n-Pintado et al.\ (1992), who derive $X({\rm SiO})
\sim 2 \times 10^{-6}$.  In HH~211, however, the SiO emission is not as
extended as the high-velocity CO jet, and so our value for $X$(SiO) is
likely to be a lower limit.  The abundance of SiO is therefore enhanced
by approximately a factor of $10^6$ over that typically observed in
quiescent molecular clouds, where $X({\rm SiO}) \la 10^{-11}$ (Ziurys,
Friberg, \& Irvine 1989).  Such a high abundance suggests that at least
10\% of all the silicon is in the form of SiO, requiring an extremely
efficient mechanism for releasing Si from dust grains into the gas phase,
to enable subsequent gas-phase chemistry to produce SiO.

Two mechanisms have been proposed for increasing the amount of gas-phase
silicon in shocks.  In J-shocks and slow C-shocks grain-grain collisions
may be the dominant means of destroying the grains to release refractory
elements (Seab \& Shull 1983; Jones et al.\ 1994; Caselli et al.\ 1997),
while in faster C-shocks the sputtering of grain mantles and refractory
cores by heavy ions is a more efficient process (Field et al.\ 1997;
Schilke et al.\ 1997; Caselli et al.\ 1997).  Once produced the SiO may
be destroyed in shocks by reactions such as SiO + C $\rightarrow$ CO +
Si, SiO + OH $\rightarrow$ SiO$_2$ + H, and SiO + $h\nu$ $\rightarrow$
SiO$^+$ + $e$ (e.g., Hartquist, Oppenheimer, \& Dalgarno 1980; Neufeld \&
Dalgarno 1989).  It may also be depleted back onto dust grains in the
post-shock gas.

Both the extremely localized nature of the SiO emission from the HH~211
jet, and the plausible explanations for producing the observed large
abundance enhancements via the destruction of dust grains in shocks,
suggest that dust must be located either within the jet itself,
or in a cocoon of material surrounding the jet and from which the
jet can entrain the dust.  However, the separation of the CO(2--1)
into two distinct kinematic components by Gueth \& Guilloteau (1999)
--- a low-velocity shell and a high-velocity, highly-collimated jet,
with no evidence for molecular material at intermediate velocities or in
the cavity between the shell and the jet --- makes it seem most likely
that the dust originates at the base of the jet, from where the dust is
entrained or injected directly into the jet.  If the limb-brightened
edges of the SiO jet are indicative of lateral entrainment the dust
must either be destroyed locally in shocks long the edges of the jet,
or the SiO must already be formed at the jet base.

The observed location and velocity of the SiO emission are consistent with
its production in C-shocks, with speeds in the range 10--50~km~s$^{-1}$.
Such shock conditions are also believed to produce most of the
observed H$_2$ emission from outflow sources (e.g., Smith, Brand, \&
Moorhouse 1991), which would lead us to expect a much better correlation
between SiO and H$_2$ emission than is observed.  Although the SiO
peaks in Fig.~\ref{sio-int} do lie on areas of strong H$_2$ emission,
beyond the SiO jet, at distances greater than $20''$ (7000~AU) from the
driving source, the H$_2$ remains strong while there is no detectable SiO
emission.  A lack of precise correspondence between SiO and H$_2$ emission
is also observed in IRAS~03282+3035 and L1157, but for these sources
the SiO emission tends to lie downstream from strong H$_2$ emission, and
each tracer is assumed to be following different physical and chemical
conditions in the evolution of the post-shock gas (Bachiller et al.\ 1994;
Gueth et al.\ 1998).  One possible explanation for the lack of detectable
SiO emission near the brightest H$_2$ peaks in HH~211 is that although
gas-phase SiO is being produced in the shocks, the post-shock density
is insufficient to collisionally excite the rotational lines; such a
picture is consistent with the SiO primarily being produced close to the
driving source, where the envelope densities are highest.  Alternatively,
it is possible that one of the two SiO destruction processes referred
to above is operating efficiently in the shocks occurring further from
the driving protostar.

\section{Conclusions}

The SiO images of the HH~211 jet presented in this paper are among the
highest resolution images of a protostellar jet: the resolution of $0.5''$
corresponds to only 170~AU, and is sufficient to resolve the width of
the jet as traced by SiO emission.  The SiO emission is extremely clumpy,
and arises in multiple compact knots whose scale is much smaller than the
coherent shocked structures traced by H$_2$ emission.  The most likely
origin for these SiO structures is in internal bowshocks generated by
a time-dependent jet velocity.  The clumpy nature of the SiO emission
suggests these shocks have a very complex spatial structure, most likely
due to Raleigh-Taylor instabilities in the cooling, post-shock gas.
There is also some morphological evidence for limb-brightened SiO
emission, which could arise in a hot boundary layer between the jet and
the surrounding material, but this pattern of emission could equally
well be generated in the tails of an internal bowshock.  Although the SiO
detected does coincide with the strong H$_2$ peaks close to the driving
protostar, no SiO emission is detected at projected distances of more than
$20''$, despite these being the locations of the strongest H$_2$ knots.
This result suggests that the shock conditions, perhaps the pre-shock
density, vary rapidly along the length of the jet.

\section*{Acknowledgments}

The authors are very grateful to Mark McCaughrean for providing the image
of shocked molecular hydrogen emission associated with HH~211, and thank
the referee for making useful comments that have helped clarify the paper.
CJC was supported by a PPARC Advanced Fellowship during part of this work,
and JSR is partially supported by a Royal Society Fellowship.

\newpage

$$\psfig{file=fig1.ps,width=6.5in,angle=0}$$

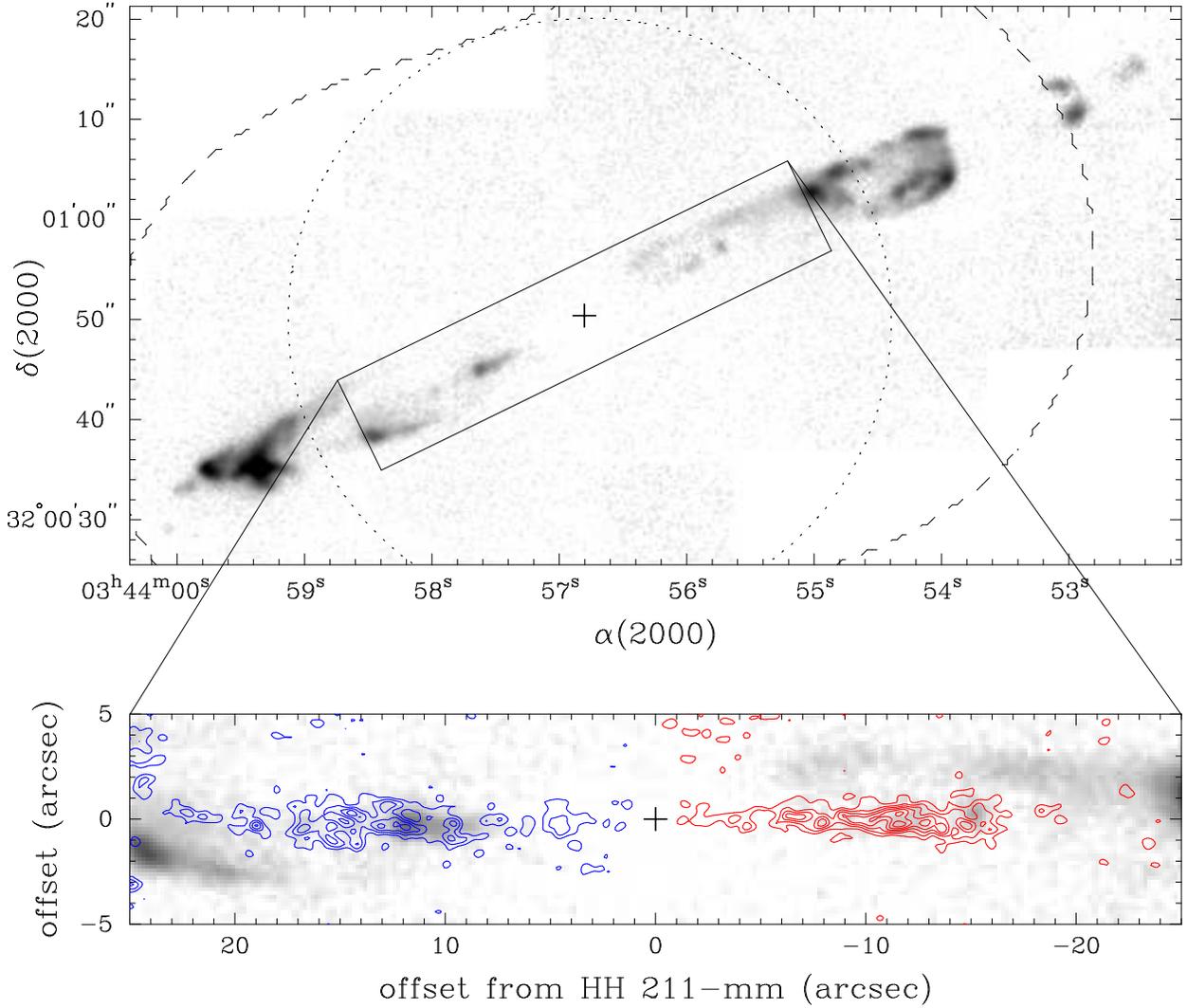
\figcaption{\label{sio-int}{\it Top:} Logarithmic greyscale image of
the near-infrared H$_2$ emission from HH~211, from McCaughrean et al.\
(1994).  The dashed contour outlines the 50\% sensitivity level of the
VLA observations obtained in D configuration, and the dotted contour
describes the same for the C-array data.  The cross marks the position
of the protostar, HH~211-mm, from Gueth \& Guilloteau (1999).  No SiO
(1--0) emission was detected in C or D configuration outside the box
(solid contour) centered on the protostar.  {\it Bottom:} Expanded image
of the box centered on HH~211-mm, rotated by $-26^\circ$ to align it with
the horizontal.  The infrared H$_2$ is in greyscale, and is overlaid by
the integrated blueshifted (blue contours, to the east of the protostar)
and redshifted (red contours, to the west) SiO(1--0) emission imaged
with the VLA\@.  Contours begin at, and are spaced at intervals of,
3~mJy~arcsec$^{-2}$.}

\newpage

$$\psfig{file=fig2.ps,height=7in,angle=0}$$

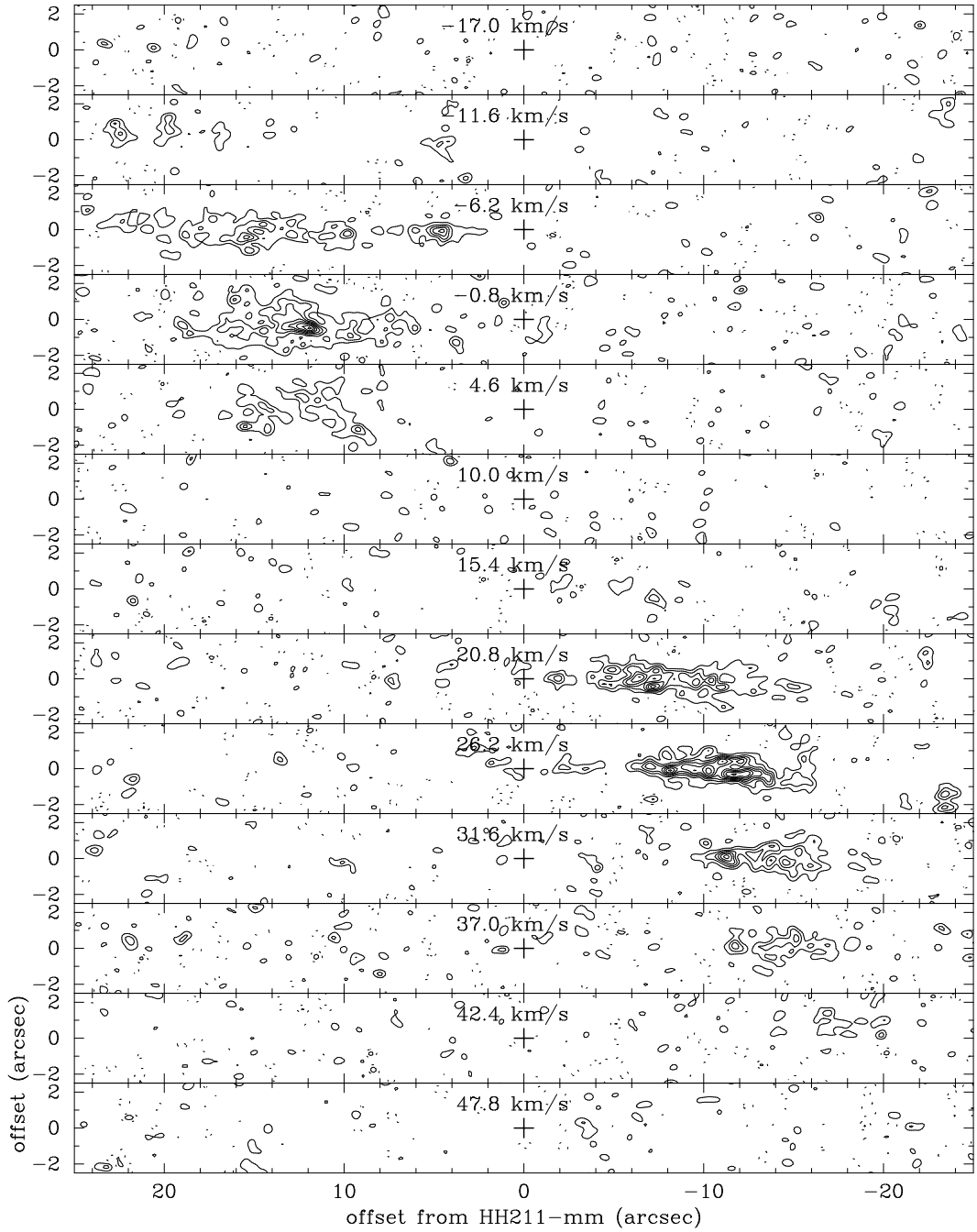
\figcaption{\label{sio-chan}Channel maps of the SiO(1--0) emission
from HH~211.  The LSR velocity is given at the top of each panel, and
contours equally spaced at intervals of 5.2~mJy~arcsec$^{-2}$.}

\newpage

$$\psfig{file=fig3.ps,width=6.5in,angle=0}$$

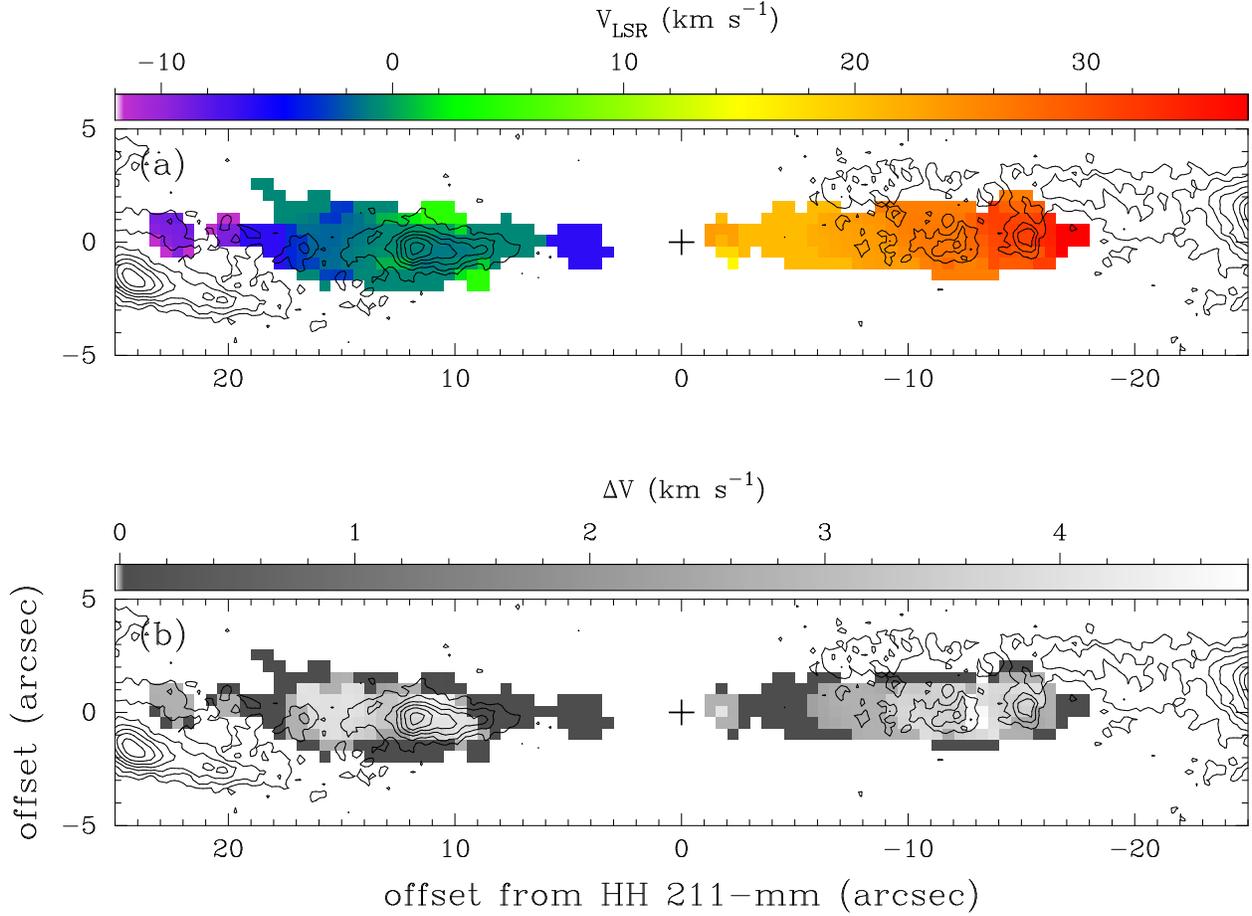
\figcaption{\label{mom-D}(a) The first moment of the SiO(1--0) emission,
obtained from the VLA D-array data only, overlaid by contours of the
infrared H$_2$ emission.  The color scale runs from $V_{\rm LSR} =
-12$~km~s$^{-1}$ (blue) to +37~km~s$^{-1}$ (red); (b) the second moment
of the SiO(1--0) emission, running from $\Delta V = 0$~km~s$^{-1}$
to 4.8~km~s$^{-1}$.  The highest velocity dispersion of the SiO(1--0)
line generally coincides with shocked H$_2$ emission.}

\newpage

\rule{0mm}{2cm}

$$\psfig{file=fig4.ps,height=6.5in,angle=-90}$$

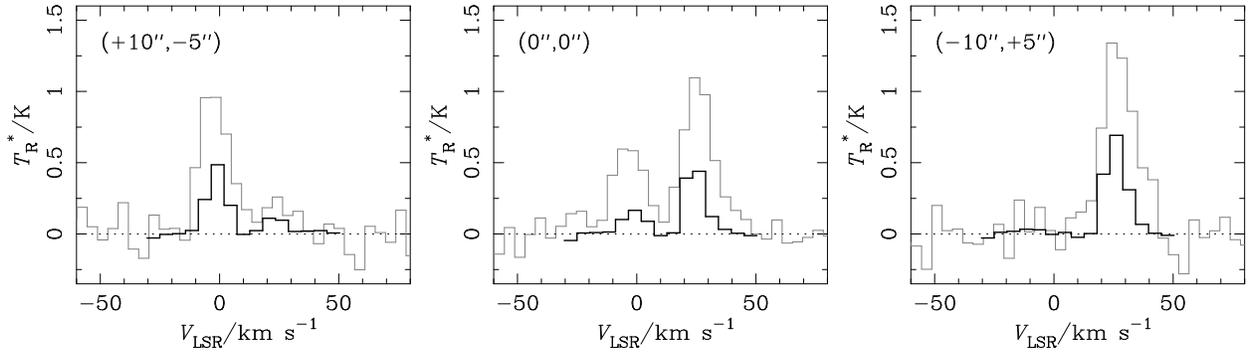
\figcaption{\label{sio-spec}Spectra of the SiO(5--4) in grey and SiO(1--0)
in black towards the blueshifted jet (left), the protostar (center),
and the redshifted jet (right).}

\newpage

\begin{deluxetable}{lcccccc}
\tabletypesize{\footnotesize}
\tablecaption{Summary of the VLA observations of SiO(1--0) in HH~211.
\label{vlatab}}
\tablehead{
\colhead{Array} & \multicolumn{2}{c}{Phase center} & Synthesised &
Restoring & \colhead{Rms} & \colhead{K~mJy$^{-1}$}\\
\colhead{} & \colhead{$\alpha(2000)$} & \colhead{$\delta(2000)$} &
\colhead{beam ($''$)} & \colhead{beam ($''$)} &
\colhead{(mJy~beam$^{-1}$)}
& \colhead{}\\
\colhead{} & \colhead{(${\rm^h~~^m~~^s}$)~~~} &
\colhead{($^\circ$~~~$'$~~~$''$)} & \colhead{} & \colhead{} &
\colhead{} & \colhead{} \\
}
\startdata
D & 03 43 58.200 & 32 00 43.00 & 2.17$\times$1.43 @ 69$^\circ$\phs &
2.17$\times$1.43 @ 69$^\circ$\phs & 4.5 & 0.21 \\
  & 03 43 55.200 & 32 00 57.00 & 2.16$\times$1.43 @ 68$^\circ$\phs &
2.17$\times$1.43 @ 69$^\circ$\phs & 4.5 & 0.21 \\
C & 03 43 56.760 & 32 00 50.00 & 0.58$\times$0.43 @ $-$44$^\circ$ &
 0.58$\times$0.43 @ $-$44$^\circ$ & 2.3 & 2.6\phn \\
\enddata
\end{deluxetable}

\end{document}